\documentclass[useAMS,usenatbib]{mn2e}

\usepackage{graphicx,amssymb,amsmath}

\title[The mass function of 35 Galactic GCs]{The global mass functions of 35
Galactic globular clusters: I. Observational data and correlations with 
cluster parameters}
\author[Sollima et al.]{A. Sollima$^{1}$\thanks{E-mail:
antonio.sollima@oabo.inaf.it}, H. Baumgardt$^{2}$\\
$^{1}$ INAF Osservatorio Astronomico di Bologna, via Gobetti 93/3, Bologna,
40129, Italy\\
$^{2}$ School of Mathematics and Physics, University of Queensland, St Lucia,
QLD 4072, Australia\\
}
\begin{document}


\pagerange{\pageref{firstpage}--\pageref{lastpage}} \pubyear{2017}

\maketitle

\label{firstpage}

\begin{abstract}
We have derived the global mass functions of a sample of 35 Galactic globular clusters
by comparing deep Hubble Space Telescope photometry with suitable multimass
dynamical models. For a subset of 29 clusters with available radial
velocity information we were also able to determine dynamical parameters,
mass-to-light ratios and the mass fraction of dark remnants. 
The derived global mass functions are well described by
single power-laws in the mass range $0.2 < m/M_\odot < 0.8$ with mass function slopes
$\alpha>-1$. Less evolved clusters show deviations from a single-power law, indicating that
the original shape of their mass distribution was not a power-law.
We find a tight anticorrelation between the present-day mass function 
slopes and the half-mass relaxation times, which can be understood if clusters started from
the same universal IMF and internal dynamical evolution is the
main driver in shaping the present-day mass functions. Alternatively, IMF
differences correlated with the present-day half-mass relaxation time are needed
to explain the observed correlation.
The large range of mass function slopes seen for our clusters implies that
most globular clusters are dynamically highly evolved, a fact that seems difficult
to reconcile with standard estimates for the dynamical evolution of clusters.
The mass function slopes also correlate with the dark remnant fractions
indicating a preferential retention of massive remnants in clusters subject to
high mass-loss rates.
\end{abstract}


\section{Introduction}
\label{intro_sec}

One of the long-standing issues of stellar astrophysics is the understanding of
the mechanisms determining the mass distribution of stars. This topic represents one of the
central questions in the theory of star formation and has strong relevance for many
areas of astrophysics. The original distribution of stellar masses, commonly 
referred to as the Initial Mass Function (IMF), is indeed a key ingredient in models of
stellar population synthesis, chemical evolution of clusters and galaxies, 
dynamical evolution of stellar systems and, in general, in any topic involving
the role of baryons. 

In this regard, the universality of the IMF, its shape and
the parameters driving its hypothetical variation are questions
still far from being completely understood from both a theoretical and an
observational point of view. Indeed, many complex processes affect the
efficiency of fragmentation of a molecular cloud (dependence of the Jeans mass
from thermodynamical parameters, competitive accretion, metal line-driven 
cooling, etc.; Silk 1977; Fleck 1982; Bonnell et al. 1997; 
Nakamura \& Umemura 2001). A practical difficulty in observationally constraining 
the IMF resides in its temporal evolution, which strongly depends on the
characteristics of the considered stellar population and on its environment.
An ideal class of astrophysical objects where to perform an analysis of the IMF 
should be young, dynamically unevolved stellar populations containing a large number of 
coeval and chemically homogeneous stars covering a wide range of masses.  
None of the known star forming complexes satisfy all the above requirements so,
from the pionering study by Salpeter (1955), many studies concentrated on the
determination of the IMF shape in the Galactic field, in OB
associations (Miller \& Scalo 1979; Kroupa 2001; Chabrier 2003) and, more
recently, in dwarf galaxies (Geha et al. 2013). Despite the huge
observational effort made during the last 60 years, there is still no clear
evidence for systematic variations of the IMF and conflicting results
have been reported in the past (see Bastian, Covey \& Meyer 2010 for a recent review).

Globular clusters (GCs) are in principle among the best places to investigate the
distribution of stellar masses at the low-mass end of the MF
($0.1<M/M_{\odot}<1$). They are composed out of hundred of thousands to millions of stars,
located at the same distance and
formed in a short time interval from a chemically relatively homogeneous cloud 
covering a wide range of masses. Moreover, there is a significant number of GCs at
distances $<$20 kpc for which it is possible to perform a statistically
meaningful sampling of their stellar population down to the hydrogen-burning
limit with a good level of completeness. On the other hand, the relaxation times
of globular clusters are often smaller than their ages so that the large
number of interactions among their stars produces a mass-dependent distribution 
of kinetic properties (energies and angular momenta).
This reflects into the time evolution of the MF since low-mass stars
progressively gain energy, being more prone to evaporation. As mass-loss
proceeds independently, the MF tends to flatten on timescales depending on both the internal
structure of the cluster and the strength of the external tidal field 
(Baumgardt \& Makino 2003 ; Lamers, Baumgardt \& Gieles 2013).
Moreover, the tendency toward energy equipartition leads to a radial segregation
of different mass groups with the most massive stars moving on less 
energetic orbits preferentially confined to the innermost cluster regions, while 
low-mass stars diffuse into an extended halo.
For all these reasons, the present-day MF measured in a particular region of a
GC does not reflect either its IMF nor its global MF.
The derivation of the present-day global MF is however still
possible by correcting the locally estimated MF by the mass-segregation
effects predicted by some suitable dynamical model (see e.g. McClure et al. 1986; Paust
et al. 2010). Such corrections depend on
the cluster concentration, MF and distance from the cluster centre but they appear to
be generally small close to the half-mass radius (Baumgardt \& Makino 2003). 
So, an alternative approach is to estimate the MF
in this region of the cluster and assume it as a good representation of the
global MF (e.g. Piotto \& Zoccali 1999).
On the theoretical side, many surveys of N-body simulations have been performed
to investigate the evolution of the MF in GC-like objects (Vesperini \& Heggie
1997; Baumgardt \& Makino
2003; Lamers, Baumgardt \& Gieles 2013; Webb \& Vesperini 2014, 2016). In
particular, Leigh et al. (2012) used a set of N-body runs assuming different
masses, concentrations, orbital eccentricities and tidal environments to
reproduce the MFs of a sample of 27 Galactic GCs and showed that the natural
evolution of a universal IMF could actually produce the observed
cluster-to-cluster differences.

Observationally, since the early 1980's many studies focussed on the determination of the MF in 
individual GCs (without correcting for incompleteness, e.g. Da Costa 1982; 
Richer et al. 1990; Santiago, Elson \& Gilmore 1996; Chabrier \& Mera
1997; Paresce \& De Marchi 2000; Pulone et al. 2003; Paust, Wilson \& van Belle
2014).
The first comprehensive studies of the MFs in a number of GCs large enough to explore
possible correlations with various cluster parameters have been those
by Capaccioli, Piotto \& Stiavelli (1993) and Djorgovski, Piotto \& Capaccioli 
(1993) who collected the MFs measured by different authors for a sample of 17 
Galactic GCs and reported a dependence of their slopes (measured using stars
with masses $m>0.5 M_{\odot}$) with the cluster position in the Galaxy.
Piotto \& Zoccali (1999) analysed in a homogeneous way deep Hubble Space 
Telescope (HST) images taken near the half-mass radii of seven globular clusters 
reaching a limiting mass of 
$m\sim0.3 M_{\odot}$. They found that the MF slopes correlate with the orbital 
destruction rates of the clusters in the Galaxy and anticorrelate with their half-mass 
relaxation times although their small sample hampered any firm conclusion on the significance 
of these correlations.
De Marchi, Paresce \& Pulone (2007) used a sample of HST and Very Large
Telescope data for a sample of 20 GCs and found a well defind correlation
between the slope of their MFs and their King model concentration parameter $c$.
Finally, Paust et al. (2010) derived the central and global present-day MFs of
17 GCs as part of the ACS Survey of Galactic Globular Clusters treasury
project (Sarajedini et al. 2007) by comparing ACS/HST photometric data with multimass dynamical models.
They found a significant correlation between the MF slope and the central 
density (or equivalently the central surface brightness), while detecting only 
marginal statistical significance of the previously reported correlations with 
other parameters.

In this paper we use the ACS treasury project database to extend the census of
GC MFs to a sample of 35 clusters, more than doubling the sample already
analysed by Paust et al. (2010). By means of a comparison with multimass 
analytical models we derive the global MFs of the analysed clusters and
investigate possible correlations with their structural  and dynamical parameters.
In Sec.~2 we present the database used in this work. The adopted
dynamical models are described in Sec.~3. Sec.~4 is devoted to the
description of the algorithm adopted
to determine global MFs and other structural parameters. The obtained MFs
and the analysis of their shapes are presented in Sec.~5. In Sec.~6 we
search for correlations with various cluster
parameters. 
We finally discuss our results in Sec.~7.

\begin{table*}
 \centering
 \begin{minipage}{200mm}
  \caption{Parameters of the best-fit models.}
  \begin{tabular}{@{}lccccccccr@{}}
  \hline
  NGC & $\alpha$ & $log (M_{lum}/M_{\odot}$) & $log (M_{dyn}/M_{\odot}$) &
  $r_{h}$ & $f_{remn}$ & log ($t_{rh}$/yr) & $M_{dyn}/L_{V}$ & $log~\rho_{0}$ & $log~\rho_{h}$\\
      &          &                           &                           &   
    pc    &            &                   & $M_{\odot}/L_{\odot}$ &
    $M_{\odot}/pc^{3}$ & $M_{\odot}/pc^{3}$\\
 \hline
288  & -0.66 $\pm$ 0.04 & 4.67 $\pm$ 0.04 & 4.96 $\pm$ 0.03 & 9.12  &  0.50 $\pm$ 0.05 &  9.60 & 1.89 $\pm$ 0.49 & 1.92 & 1.16\\
1261 & -0.65 $\pm$ 0.03 & 4.93 $\pm$ 0.10 & 5.23 $\pm$ 0.05 & 5.70  &  0.50 $\pm$ 0.11 &  9.39 & 1.51 $\pm$ 0.53 & 3.26 & 2.04\\
1851 & -0.69 $\pm$ 0.03 & 5.14 $\pm$ 0.07 & 5.51 $\pm$ 0.04 & 5.14  &  0.57 $\pm$ 0.08 &  9.43 & 1.64 $\pm$ 0.49 & 4.49 & 2.45\\
2298 &  0.11 $\pm$ 0.03 & 4.18 $\pm$ 0.12 &	            & 3.19  &                  &       &        	 &	& \\
3201 & -1.26 $\pm$ 0.09 & 4.81 $\pm$ 0.03 & 5.08 $\pm$ 0.03 & 6.41  &  0.47 $\pm$ 0.05 &  9.46 & 1.95 $\pm$ 0.50 & 3.05 & 1.74\\
4147 &  0.03 $\pm$ 0.05 & 4.21 $\pm$ 0.13 & 4.81 $\pm$ 0.22 & 5.22  &  0.75 $\pm$ 0.26 &  9.12 & 2.13 $\pm$ 1.36 & 3.83 & 1.73\\
4590 & -1.27 $\pm$ 0.07 & 4.87 $\pm$ 0.06 & 5.29 $\pm$ 0.06 & 8.60  &  0.62 $\pm$ 0.09 &  9.74 & 2.91 $\pm$ 0.89 & 3.12 & 1.56\\
4833 & -0.69 $\pm$ 0.08 & 5.10 $\pm$ 0.03 & 5.35 $\pm$ 0.07 & 8.60  &  0.43 $\pm$ 0.08 &  9.71 & 1.32 $\pm$ 0.39 & 3.21 & 1.62\\
5024 & -1.41 $\pm$ 0.11 & 5.53 $\pm$ 0.05 & 5.82 $\pm$ 0.06 & 15.12 &  0.49 $\pm$ 0.08 & 10.34 & 1.95 $\pm$ 0.56 & 3.09 & 1.36\\
5053 & -1.29 $\pm$ 0.03 & 4.62 $\pm$ 0.03 & 4.78 $\pm$ 0.11 & 19.30 &  0.32 $\pm$ 0.11 & 10.11 & 1.64 $\pm$ 0.57 & 0.59 & 0.00\\
5272 & -0.95 $\pm$ 0.08 & 5.35 $\pm$ 0.05 & 5.61 $\pm$ 0.04 & 7.29  &  0.45 $\pm$ 0.07 &  9.73 & 1.68 $\pm$ 0.46 & 3.76 & 2.10\\
5286 & -0.61 $\pm$ 0.03 & 5.29 $\pm$ 0.07 & 5.71 $\pm$ 0.08 & 4.43  &  0.61 $\pm$ 0.11 &  9.43 & 1.30 $\pm$ 0.45 & 4.28 & 2.85\\
5466 & -1.68 $\pm$ 0.09 & 4.82 $\pm$ 0.03 & 4.77 $\pm$ 0.11 & 24.64 & -0.12 $\pm$ 0.12 & 10.25 & 1.25 $\pm$ 0.45 & 0.98 & -0.33\\
5904 & -0.88 $\pm$ 0.10 & 5.27 $\pm$ 0.04 & 5.58 $\pm$ 0.04 & 7.66  &  0.51 $\pm$ 0.06 &  9.73 & 1.99 $\pm$ 0.53 & 3.86 & 2.00\\
5986 & -0.65 $\pm$ 0.07 & 5.20 $\pm$ 0.07 & 5.48 $\pm$ 0.05 & 5.47  &  0.48 $\pm$ 0.09 &  9.46 & 1.43 $\pm$ 0.44 & 3.53 & 2.34\\
6093 & -0.14 $\pm$ 0.04 & 5.10 $\pm$ 0.08 & 5.58 $\pm$ 0.07 & 3.39  &  0.67 $\pm$ 0.11 &  9.16 & 1.54 $\pm$ 0.52 & 4.97 & 3.07\\
6101 & -1.60 $\pm$ 0.15 & 5.11 $\pm$ 0.03 &	            & 18.75 &                  &       &        	 &	& \\
6144 & -0.15 $\pm$ 0.06 & 4.42 $\pm$ 0.06 &	            & 5.82  &                  &       &        	 &	& \\
6205 & -0.60 $\pm$ 0.08 & 5.34 $\pm$ 0.04 & 5.77 $\pm$ 0.03 & 6.87  &  0.63 $\pm$ 0.06 &  9.71 & 2.01 $\pm$ 0.53 & 3.43 & 2.34\\
6218 & -0.32 $\pm$ 0.04 & 4.60 $\pm$ 0.06 & 4.95 $\pm$ 0.03 & 4.09  &  0.55 $\pm$ 0.06 &  9.01 & 1.50 $\pm$ 0.41 & 3.38 & 2.19\\
6254 & -0.48 $\pm$ 0.09 & 4.94 $\pm$ 0.05 & 5.33 $\pm$ 0.05 & 5.21  &  0.59 $\pm$ 0.07 &  9.34 & 1.93 $\pm$ 0.55 & 3.79 & 2.26\\
6304 & -1.89 $\pm$ 0.19 & 5.17 $\pm$ 0.05 & 5.27 $\pm$ 0.05 & 5.69  &  0.20 $\pm$ 0.07 &  9.57 & 3.05 $\pm$ 0.86 & 4.13 & 2.08\\
6341 & -0.75 $\pm$ 0.05 & 5.15 $\pm$ 0.06 & 5.48 $\pm$ 0.03 & 5.39  &  0.53 $\pm$ 0.07 &  9.47 & 1.56 $\pm$ 0.44 & 4.47 & 2.36\\
6397 & -0.40 $\pm$ 0.03 & 4.61 $\pm$ 0.02 & 4.96 $\pm$ 0.02 & 4.60  &  0.56 $\pm$ 0.03 &  9.14 & 1.09 $\pm$ 0.26 & 5.65 & 2.05\\
6541 & -0.49 $\pm$ 0.05 & 5.09 $\pm$ 0.06 & 5.41 $\pm$ 0.05 & 4.64  &  0.53 $\pm$ 0.08 &  9.32 & 1.64 $\pm$ 0.49 & 4.85 & 2.49\\
6584 & -0.53 $\pm$ 0.02 & 4.62 $\pm$ 0.13 &	            & 4.67  &                  &       &        	 &	& \\
6656 & -0.98 $\pm$ 0.13 & 5.42 $\pm$ 0.02 & 5.69 $\pm$ 0.03 & 6.25  &  0.46 $\pm$ 0.04 &  9.70 & 1.86 $\pm$ 0.46 & 3.88 & 2.38\\
6723 & -0.24 $\pm$ 0.05 & 4.84 $\pm$ 0.07 & 5.23 $\pm$ 0.11 & 5.04  &  0.59 $\pm$ 0.13 &  9.28 & 1.91 $\pm$ 0.71 & 3.37 & 2.20\\
6752 & -0.49 $\pm$ 0.07 & 4.97 $\pm$ 0.03 & 5.38 $\pm$ 0.02 & 5.68  &  0.60 $\pm$ 0.04 &  9.44 & 1.94 $\pm$ 0.48 & 5.03 & 2.19\\
6779 & -0.55 $\pm$ 0.03 & 4.79 $\pm$ 0.09 &	            & 4.92  &                  &       &        	 &	& \\
6809 & -0.89 $\pm$ 0.05 & 4.90 $\pm$ 0.03 & 5.29 $\pm$ 0.03 & 6.31  &  0.59 $\pm$ 0.04 &  9.50 & 1.83 $\pm$ 0.45 & 2.81 & 1.97\\
6934 & -0.77 $\pm$ 0.04 & 4.84 $\pm$ 0.10 &	            & 5.98  &                  &       &        	 &	& \\
7078 & -1.16 $\pm$ 0.07 & 5.54 $\pm$ 0.05 & 5.81 $\pm$ 0.03 & 7.71  &  0.47 $\pm$ 0.06 &  9.89 & 1.79 $\pm$ 0.48 & 4.16 & 2.23\\
7089 & -0.83 $\pm$ 0.07 & 5.53 $\pm$ 0.05 & 5.89 $\pm$ 0.06 & 7.87  &  0.56 $\pm$ 0.08 &  9.88 & 1.98 $\pm$ 0.57 & 4.11 & 2.28\\
7099 & -0.72 $\pm$ 0.02 & 4.82 $\pm$ 0.07 & 5.16 $\pm$ 0.05 & 5.53  &  0.54 $\pm$ 0.08 &  9.37 & 1.48 $\pm$ 0.44 & 5.04 & 2.01\\
\hline
\end{tabular}
\end{minipage}
\end{table*}

\section{Observational material}
\label{obs_sec}

The derivation of the global MFs and the cluster parameters has been 
performed through the analysis of three different kinds of datasets: photometry,
surface brightness profiles and individual stellar radial velocities.

The photometric data consists of high-resolution HST observations of a 
sample of 66 Galactic GCs obtained as part of the {\it Globular Cluster ACS 
Treasury Project} (Sarajedini et al. 2007). The data has been obtained using deep images 
obtained with the Advanced Camera for Surveys (ACS) Wide Field Channel through 
the F606W and F814W filters. The field of view of the camera 
($202" \times 202"$) was centered on the cluster centres with a dithering 
pattern to cover the gap between the two chips, allowing a full coverage of the 
core of all the GCs considered in our analysis. This survey provides deep 
colour-magnitude diagrams (CMDs) providing photometry of main sequence stars
down to the hydrogen burning limit (at $M_{V}\sim 10.7$) with a signal-to-noise
ratio $S/N > 10$ for all target clusters. The results of artificial star experiments 
are also available to allow an accurate estimate of the completeness level and 
photometric errors. A detailed description of the photometric reduction, 
astrometry, and artificial star experiments can be found in Anderson et al. 
(2008). Within this database we excluded from our analysis all GCs with
{\it i)} evidence of large ($\Delta Y>0.1$) helium variation ($\omega$ Cen, 
NGC 2808, NGC 6388, NGC 6441), {\it ii)} significant
contamination by either bulge (NGC 6624, NGC 6637) or Sagittarius dwarf galaxy
stars (M54), or {\it iii)} a completeness level estimated in the innermost
arcminute at the hydrogen-burning limit smaller than $\psi<10\%$. 
Thirty-five GCs passed the above selection criteria (see Table 1).

The surface brightness profiles for most GCs of our sample were taken from
Trager, King \& Djorgovski
(1995). They were constructed from generally inhomogeneous data based 
mainly on the Berkeley Globular Cluster Survey (Djorgovski \& King 1984). 
The surface brightness profile of each cluster has been derived by matching 
several sets of data obtained with different techniques (aperture photometry on 
CCD images and photographic plates, photoelectric observations, star counts, 
etc.). Moreover, the profiles of the more distant and/or faint GCs are often
noisy and do not extend beyond a few core radii. 
For this reason we adopted, where available, the number density profiles 
calculated by Miocchi et al. (2013) from wide field photometry. Finally, the
density profile calculated by Melbourne et al. (2000) and Alonso-Garc{\'{\i}}a et al.
(2012) have been adopted for
NGC 4833 and NGC 6144, respectively. Because of the better angular resolution of HST data, ACS observations
sample the innermost portion of our clusters much more accurately
than any other previous ground based analysis. For this reason, the surface
brightness profile of the innermost $1.6\arcmin$ has been calculated directly
from ACS data by summing completeness-corrected F606W fluxes 
$$\mu=-2.5 log \left(\sum_{i} \frac{10^{-0.4 F606W_{i}}}{c_{i}} \right)$$ 
in annuli of $0.1\arcmin$ width and matched to the adopted external profile using the overlap
region. The completeness factors $c_{i}$ have been calculated for all stars as
the fraction of recovered objects\footnote{An artificial star has been
considered recovered if its input and output magnitudes differ by less than 2.5
log(2) ($\sim0.75$) mag in both F606W and F814W magnitudes.} in the artificial star catalog among
all stars within $0.\arcmin05$ from the position of each individual star and within
$0.25$ mag of the F606W and F814W magnitudes of each individual star.

Among the 35 GCs of our sample we found large sets ($>$50) of available radial 
velocities in the literature for 29 of them (see Table A1 of Baumgardt 2017 for 
the references for each cluster). Radial velocities from different sources were corrected for
systematic shifts using the stars in common. 
Additional radial velocities for clusters NGC 1261, NGC 5986, NGC 6304 and NGC 6541 
were derived from archival FLAMES@VLT spectra collected under the observing programmes
193.D-0232 (PI Ferraro) and 093.D-0628 (PI Zocchi).
For this task, pipeline-reduced spectra were cross-correlated with the solar
spectrum observed with the same setups as the science observations using the task
{\it fxcor} within the {\rm IRAF} package\footnote{IRAF is distributed by the National Optical Astronomy Observatories,
which are operated by the Association of Universities for Research
in Astronomy, Inc., under cooperative agreement with the National
Science Foundation.}. 

\section{models}
\label{mod_sec}

As explained in Sec.~\ref{intro_sec}, in dynamically evolved stellar systems
like GCs, the distribution of stars depends on their mass. Hence, in order 
to derive the global MF of our target clusters, their MFs 
measured in the ACS field of view need to be corrected using the prescriptions of a dynamical model.
The structure and kinematics of our clusters have been modelled
with a set of isotropic
multimass King-Michie models (Gunn \& Griffin 1979). According to this model, 
the distribution function is given by the sum of the contribution of several
mass groups

\begin{eqnarray}
\label{eq_df}
f(E,L)&=&\sum_{j=1}^{H} k_{j} 
\left[ exp\left(-\frac{A_{j}E}{\sigma_{K}^{2}}\right)-1 \right]\nonumber\\     
\sum_{j=1}^{H}f_{j}(m_{j},r,v)&=&\sum_{i=1}^{H}k_{j}
\left[exp\left(-\frac{A_{j}(v^{2}+2\psi)}{2\sigma_{K}^{2}}\right)-1
\right]
\end{eqnarray}

where E is the energy per unit mass, $m_{j}$ is the mass of the stars in the 
j-th component, H is the number
of mass components, $k_{j}$ are coefficients determining the relative fraction
of stars in the j-th mass group, $\sigma_{K}^{2}$ is an energy normalization
constant, $r$ and $v$ are the 3D distance from the cluster centre and velocity, and 
$A_{i}$ are coefficients governing the kinetic energy balance among different
mass groups. In the
original formulation by Gunn \& Griffin (1979) $A_{j}\propto m_{j}$. Although
this last assumption is arbitrary, it has been shown that it reproduces the
structure and the degree of mass segregation of both N-body simulations during
most of their evolution and real GCs (Sollima et al. 2015, 2017).
In principle, a degree of radial anisotropy can be included by multiplying the 
distribution function of eq. \ref{eq_df} by a term dependent on the angular 
momentum. However, because of the lack of accurate proper
motion determinations for the GCs analysed here, no stringent constraints 
on the degree of anisotropy can be put. Moreover, the recent analysis by Watkins et al. (2015) based on accurate HST proper motions in 22
Galactic GCs showed that they appear to be isotropic across the field of view analysed by these authors. 
For the above reasons, we prefer to consider only isotropic
models to limit the number of free parameters.
We adopted 23 mass groups: 8 evenly spaced bins comprised between 0.1
$M_{\odot}$ and the mass at the tip of the Red Giant Branch ($M_{tip}$) and 15 
evenly spaced bins between $M_{tip}$ and 2.6 $M_{\odot}$ (i.e. the largest mass
allowed in our synthetic population; see Sec.~\ref{met_sec}).

The distribution function in eq. \ref{eq_df} can be integrated over the velocity
domain to obtain the 3D density and velocity dispersion of each mass group.

\begin{eqnarray}
\label{eq_den}
\nu_{j}(r)&=&\int_{0}^{\sqrt{-2\psi(r)}}4\pi v^{2} k_{j} f_{j}(v,r,m_{j}) dv\nonumber\\
\sigma_{v,j}^{2}(r)&=&\frac{\int_{0}^{\sqrt{-2\psi(r)}}4\pi k_{j} v^{4} f_{j}(v,r,m_{j})
dv}{\nu_{j}(r)}
\end{eqnarray}

while the potential at any radius is determined by the Poisson equation  

\begin{equation}
\label{poiss_eq}
\nabla^{2} \psi=4 \pi G \rho
\end{equation}

where

$$\rho(r)=\sum_{j=1}^{H} m_{j} \nu_{j}(r)$$

Equations \ref{eq_df} to \ref{poiss_eq} have been integrated after assuming, 
as a boundary condition, a value of the potential and its derivative at the 
centre ($\psi_{0}$; $d\psi/dr(0)=0$) outward till the radius $r_t$ at which 
both density and potential vanish.

Observational quantities (global MF, surface brightness and line-of-sight velocity
dispersion profiles) can be obtained through the relations

\begin{eqnarray}
\label{eq_obs}
N_{j}&=&4 \pi \int_{0}^{r_{t}} r^{2} \nu_{j}~dr\nonumber\\
\mu(R)&=&-2.5 log \left( \sum_{j=1}^{H}\Gamma_{j} \Upsilon_{j}\right)\nonumber\\
\sigma_{LOS,j}^{2}(R)&=&\frac{2}{3 \Gamma_{j}(R)}\int_{R}^{r_{t}}\frac{\nu_{j}
\sigma_{v,j}^{2} r}{\sqrt{r^{2}-R^{2}}}~dr
\end{eqnarray}

where

$$\Gamma_{j}(R)=2\int_{R}^{r_{t}}\frac{\nu_{j} r}{\sqrt{r^{2}-R^{2}}}~dr$$
is the projected number density and 
$\Upsilon_{j}$ is the average V-band flux of 
stars in the j-th component.
 
These models are completely defined by the free parameters
$W_{0}\equiv -\psi_{0}/\sigma_{K}^{2}$, $k_{j}$ (unequivocally defining the shape
of all profiles), $r_{c}\equiv\sqrt{9 \sigma_{K}^{2}/4 \pi G \rho(0)}$ (defining
the size of the model), and $\sigma_{K}^{2}$
(determining the normalization in mass and velocity dispersion).
The total mass and luminosity of the model can finally be calculated as

\begin{eqnarray*}
\label{eq_ml}
M&=&\sum_{j=1}^{H} N_{j} m_{j}\nonumber\\
L_{V}&=&\sum_{j=1}^{H} N_{j} \Upsilon_{j}\nonumber
\end{eqnarray*}

\section{Method}
\label{met_sec}
 
\begin{figure}
 \includegraphics[width=8.6cm]{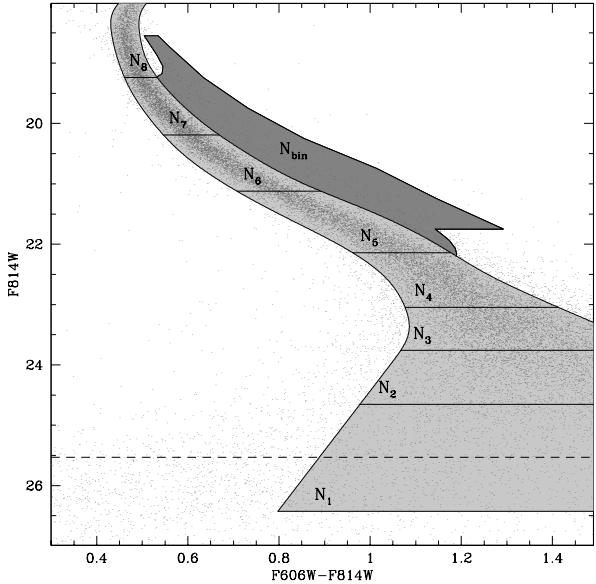}
 \caption{Selection boxes adopted for the population of single stars ($m_{1}$ to
 $m_{8}$) and binaries (bin) of NGC 288. The observed CMD 
 is overplotted. The 50\% completeness limit is marked by the dashed line.}
\label{box}
\end{figure}
  
\begin{figure*}
 \includegraphics[width=14cm]{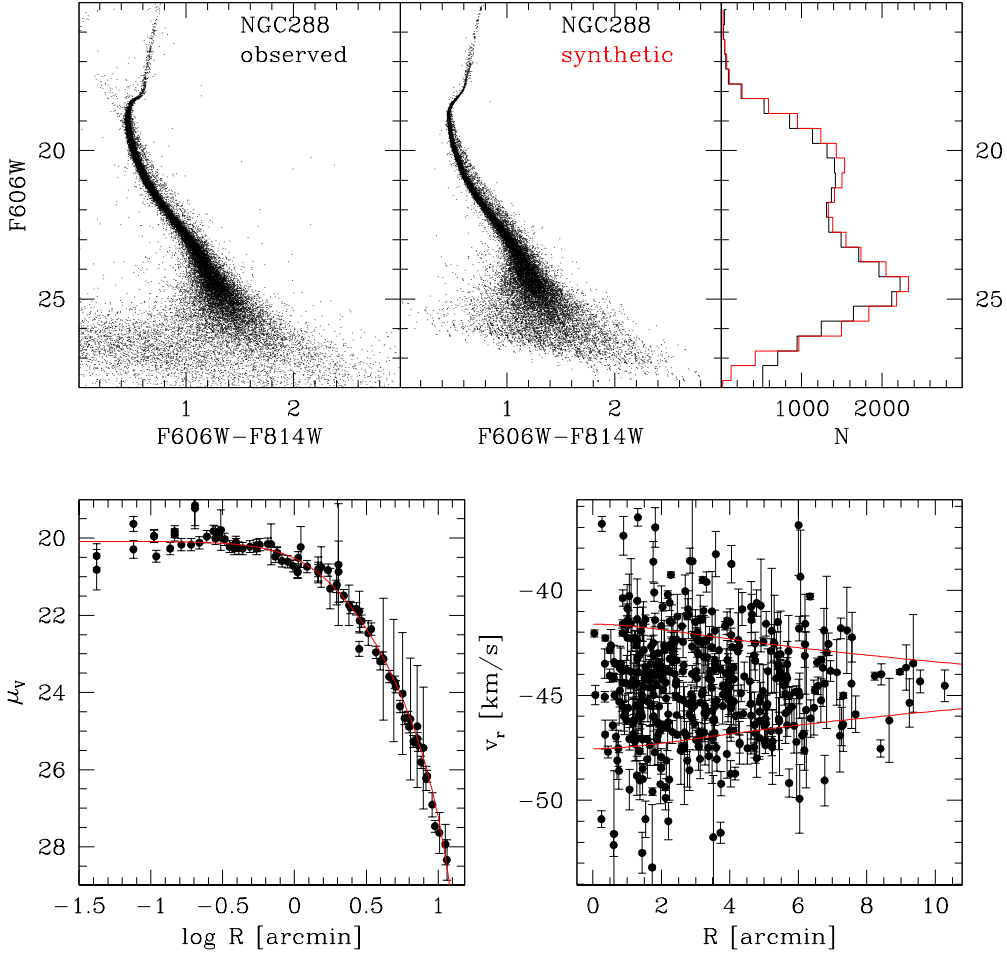}
 \caption{Comparison between three observables of NGC 288 and the corresponding 
 model prediction (red lines; grey in the printed version of the paper). Bottom-left panels: surface brightness profile; 
 bottom-right panel: velocity dispersion profile; upper-left panels: CMDs;
 upper-right panels: F606W luminosity function. The predicted 
 $\pm\sigma_{LOS}$ range is indicated in the bottom-right panel. 
}
\label{fit}
\end{figure*}

The algorithm adopted to determine the global MF of each GC is similar to that
described in Sollima, Bellazzini \& Lee (2012) and Sollima et al. (2017) and can
be schematically described as follows:
\begin{enumerate}
\item{As a first step, a synthetic stellar population has been constructed by randomly
extracting $N=10^{6}$ stars from a Kroupa (2001) IMF between 0.1 and 8
$M_{\odot}$. A fraction $f_{b}$ of binaries has also been simulated by 
randomly pairing $N_{b}=2 N f_{b}$ synthetic stars. All single stars and stars
in binaries with masses $m>M_{tip}$ have been
turned into compact remnants following the prescription
$$m_{WD}=0.109 m + 0.428~~~\mbox{(Kalirai et al. 2009)}$$
Due to the adopted upper limit of the IMF, only white dwarfs are created in this process.
This is consistent with the assumption that all neutron stars and black holes
are ejected in the early stage of cluster evolution because of the
effect of natal kicks and/or Spitzer instability (Krujissen 2009).}
\item{The corresponding synthetic CMD has been constructed by interpolating the masses
of visible stars with the mass-luminosity relation of a suitable isochrone from the Dotter et 
al. (2007) database.
For each cluster, the isochrone metallicity, age and $\alpha$-enhancement as
well as the reddening and distance modulus listed in
Dotter et al. (2010), providing an excellent fit to the ACS data, have been adopted
A synthetic horizontal branch (HB) has been simulated for each cluster using 
the tracks by Dotter et al. (2007), tuning the mean mass and mass dispersion 
along the HB to reproduce the observed HB morphology. The magnitude and colours
of binary systems have been calculated by summing the fluxes of the two
components in both passbands. We do not account for the negligible contamination
of fore/background
Galactic field interlopers possibly present in the ACS field of view. Indeed, the
Galactic model of Robin et al. (2003) predicts $<$50 field stars, corresponding to a
fraction $<$0.1\%, contained in the selection box (see below) and within the innermost
1.6\arcmin even in the low-latitude GCs of our sample.} 
\item{The synthetic stars (singles, binaries and remnants) have been
binned in mass (see Sec.~\ref{mod_sec}) and for each bin a fraction $1-X_{j}$ 
of particles has been randomly rejected. The average F606W fluxes ($\Upsilon_{j}$) of 
the remaining stars in each mass group have also been calculated;} 
\item{A dynamical model is constructed tuning the parameters $W_{0}$ and $r_{c}$ 
in order to obtain the best fit of the surface brightness profile and the $k_{j}$ coefficients
are modified to reproduce the MF of the mock catalog (see eq. \ref{eq_obs});}
\item{The distribution in phase-space (3D position and velocity) of 
synthetic stars was then extracted from the modelled distribution function using
the von Neumann rejection technique: for each star a random position in phase-space
$(r,v)$ is extracted and assigned to the star only if a
random number between 0 and 1 turns out to be smaller than
$f(m_{i},r,v)/f(m_{i},0,0)$. Projected distance and LOS velocities have then
been calculated assuming an isotropic distribution;} 
\item{For each synthetic star, a particle from the artificial star catalog 
with distance from the cluster centre within $0.\arcmin05$ and magnitudes within
0.25 mag with respect to the same quantities of the given star, has been extracted and, if
recovered, the magnitude and 
colour shift with respect to its input quantities 
have been added to those of the corresponding star. In this way, a mock CMD
accounting for photometric errors and incompleteness has been obtained;}
\item{The number of stars within $1.\arcmin6$ from the cluster centre (i.e. the
extent of the ACS field of view) and contained in 9 regions of both the observed
and the mock CMD have been counted (see Fig. \ref{box}). In particular, we defined
\begin{itemize}
\item{eight F606W magnitude intervals corresponding to 
the first 8 mass intervals and including all stars with colours 
within three times the photometric error corresponding to their magnitudes;} 
\item{a region including the bulk of the binary population with high mass ratios
($q>0.5$). This last region is delimited in magnitude by the loci of binaries with primary 
star mass $m_{1} = 0.45 M_{\odot}$ (faint boundary) and $m_{1} = 0.75
M_{\odot}$ (bright boundary), and in colour by the MS ridge line (blue boundary) 
and the equal-mass binary sequence (red boundary), both redshifted by three 
times the photometric error.}
\end{itemize}}
\item{The retention fractions ($X_{j}$) of stars in the eight mass bins and the
global binary fraction $f_{b}$ are adjusted 
by multiplying them for corrective terms which are proportional to the ratio 
between the relative number counts in each bin of the observed sample and the
corresponding model prediction
\begin{eqnarray*}
X_{j}'&=&X_{j}\left(\frac{Q_{j}^{obs} Q_{8}^{mock}}{Q_{8}^{obs} Q_{j}^{mock}}\right)^{\eta}\nonumber\\
f_{b}'&=&f_{b}\left(\frac{Q_{bin}^{obs}}{Q_{bin}^{mock}}
\frac{\sum_{j=1}^{8} Q_{j}^{mock}}
{\sum_{j=1}^{8} Q_{j}^{obs}}
\right)^{\eta}\nonumber
\end{eqnarray*}
where $Q_{j}$ and $Q_{bin}$ are the number of stars observed in the j-th single 
and in the binary selection boxes respectively, the superscripts $obs$ and
$mock$ indicate counts measured either in the observed or in the mock catalog 
respectively, and
$\eta$ is a softening parameter, set to 0.5, used to avoid divergence.
All the coefficients $X_{j}$ with $j>8$ have been set equal to 1.
}
\end{enumerate}
Steps from {\it (iii)} to {\it (viii)} have been repeated until convengerce.

For the first step we adopted $X_{j}=1$ for all j and $f_{b}=5\%$.
The above described procedure converged after $\sim20$ iterations for all the
considered clusters.
The global MF of single stars can therefore be calculated directly from the
mock catalog simulated in the last iteration. 

A final step is constituted by the mass normalization of the model. This can be done
by best fitting two independent quantities: {\it i)} the actual number of stars
in the observed ($Q^{obs}$) and in the mock ($Q^{mock}$) CMDs, 
and {\it ii)} the amplitude of the velocity dispersion
profile. The former way allows to estimate the mass in luminous stars as
$$M_{lum}=\frac{Q^{obs}}{Q^{mock}}\sum_{i=1}^{N_{sin}+N_{bin}} m_{i}$$
where the sum is extended to all single and binary stars in the final mock 
catalog excluding remnants.
For clusters with available radial velocities, the latter way provides an 
estimate of the dynamical mass ($M_{dyn}$). 
The best fit value of $M_{dyn}$ has been chosen as the one minimizing the
penalty function

$$\mathcal{L}=\sum_{i=1}^{N} \left(\frac{(v_{i}-\overline{v})^{2}}{\sigma_{LOS,8}^{2}(R_{i})+
\epsilon_{i}^{2}}+ln(\sigma_{LOS,8}^{2}(R_{i})+\epsilon_{i}^{2})\right)$$

where $v_{i}$ is the radial velocity of the i-th star, $\overline{v}$ is the systemic
velocity of the cluster, $\epsilon_{i}$ is the individual uncertainty on the radial
velocity and $\sigma_{LOS,8} (R_{i})$ is the line-of-sight velocity dispersion
predicted by the best fit model at the projected distance $R_{i}$ of the i-th
star for the 8-th mass group. The choice of
the 8-th mass bin is because radial velocities are available only for
stars along the Red Giant Branch which cover a restricted range of masses. 
Because of the dependence of kinematics on mass, it is therefore necessary to
compare the observed velocity dispersion profile with that of the corresponding
mass bin to avoid bias in the mass estimate.

Once luminous and dynamical masses are determined the fraction of dark mass can
be calculated as 
$$f_{remn}=1-\frac{M_{lum}}{M_{dyn}}$$

Finally, the central density $\rho_{0}$, the half-mass radius $r_{h}$, the
$M_{dyn}/L_{V}$ ratio of the
best fit model are computed as well as the half-mass 
relaxation time as
\begin{equation}
\label{trh_eq}
t_{rh}=0.138\frac{M_{dyn}^{1/2} r_{h}^{3/2}}{G^{1/2} \overline{m}~ln (\gamma~
M_{dyn}/\overline{m})}~~~\mbox{(Spitzer 1987)}
\end{equation}

with $\gamma=0.11$ (Giersz \& Heggie 1996). The outcome of the application of the above thechnique for NGC 288 is shown in 
Fig. \ref{fit}, as an example.

\section{results}
\label{res_sec}

\begin{figure}
\includegraphics[width=8.6cm]{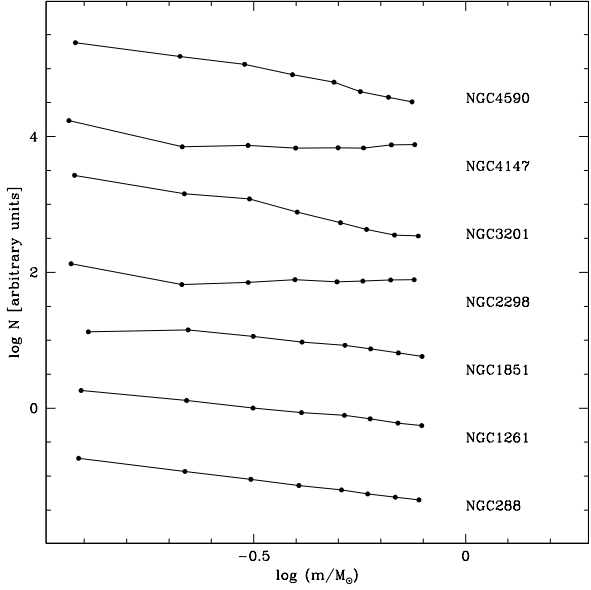}
\caption{Global MFs of NGC 288, NGC 1261, NGC 1851, NGC 2298, NGC 3201, NGC 4147
and NGC 4590. An arbitrary shift has
been added to each MF for clarity.}
\label{sho1}
\end{figure}

\begin{figure}
\includegraphics[width=8.6cm]{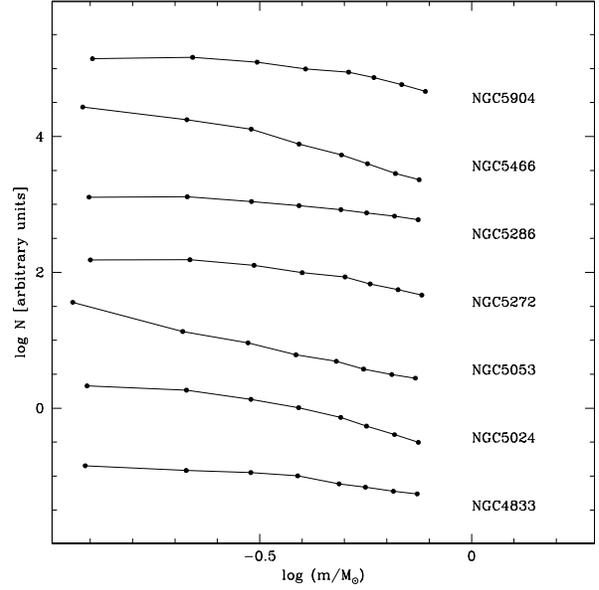}
\caption{Same as Fig. \ref{sho1} but for NGC 4833, NGC 5024, NGC 5053, NGC 5272,
NGC 5286, NGC 5466 and NGC 5904.}
\label{sho2}
\end{figure}

\begin{figure}
\includegraphics[width=8.6cm]{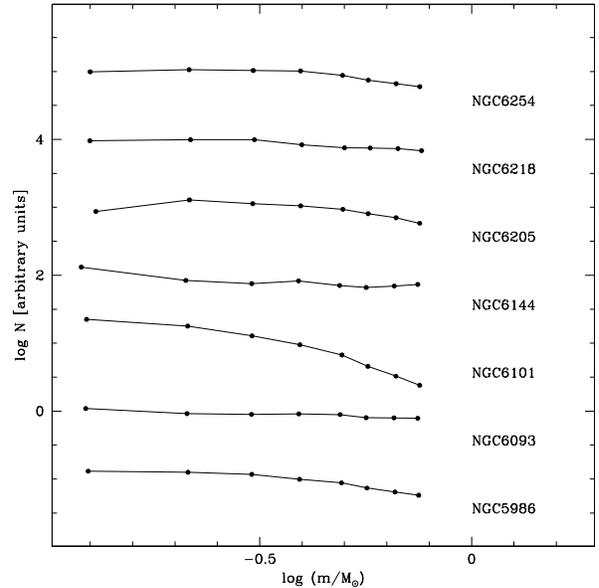}
\caption{Same as Fig. \ref{sho1} but for NGC 5986, NGC 6093, NGC 6101, NGC 6144,
NGC 6205, NGC 6218 and NGC 6254.}
\label{sho3}
\end{figure}

\begin{figure}
\includegraphics[width=8.6cm]{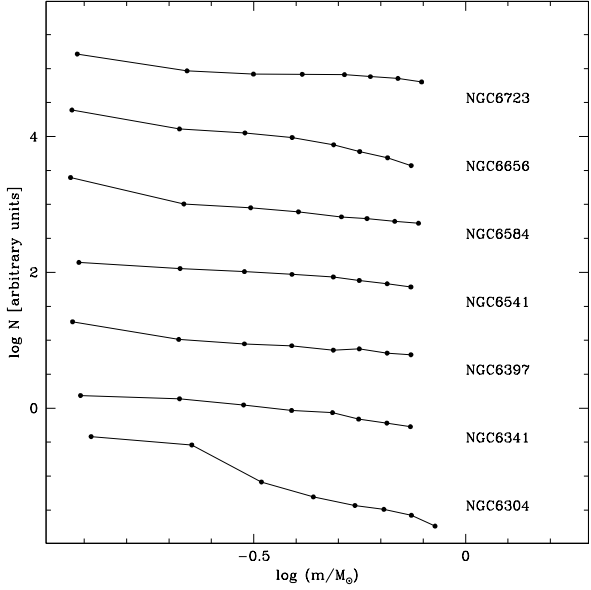}
\caption{Same as Fig. \ref{sho1} but for NGC 6304, NGC 6341, NGC 6397, NGC 6541,
NGC 6584, NGC 6656 and NGC 6723.}
\label{sho4}
\end{figure}

\begin{figure}
\includegraphics[width=8.6cm]{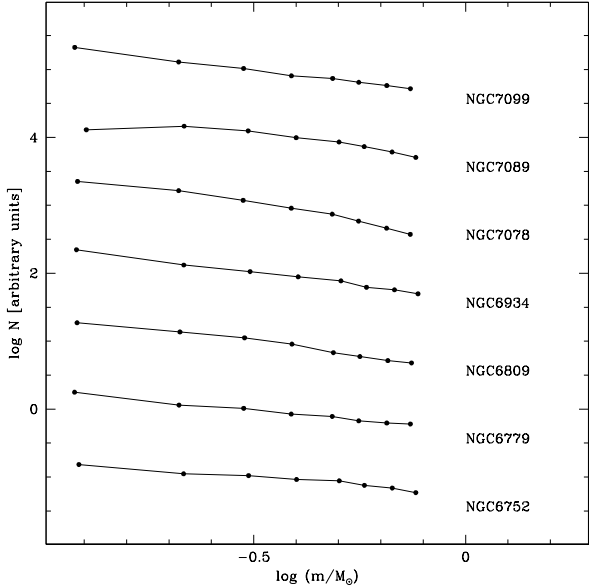}
\caption{Same as Fig. \ref{sho1} but for NGC 6752, NGC 6779, NGC 6809, NGC 6934,
NGC 7078, NGC 7089 and NGC 7099.}
\label{sho5}
\end{figure}

\begin{figure}
 \includegraphics[width=8.6cm]{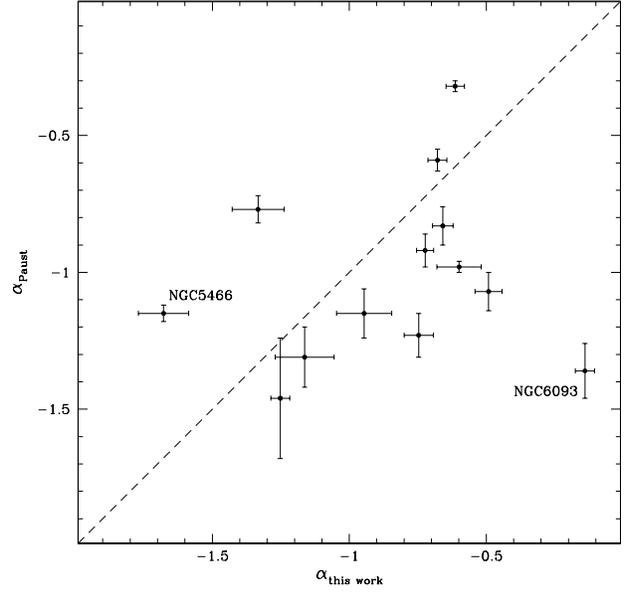}
 \caption{Comparison between the MF slopes derived in this work and those
 determined by Paust et al. (2010) for the 13 GCs in common. The one-to-one
 relation is marked by the dashed line. The location of NGC5466 and NGC6093 is
 shown.}
\label{comp}
\end{figure}

The global MFs of the 35 GCs analysed in this work are shown in Figs. \ref{sho1}
to \ref{sho5}
and the derived dynamical parameters are listed in Table 1.
Among the various parameters, the power-law index $\alpha$ of the MF has been 
calculated for stars more massive than 0.2 $M_{\odot}$ since stars below 
this limit often show relatively low levels of completeness ($\psi<50\%$)
and their relative fraction is subject to large errors. For testing purpose, 
we also calculate $\alpha$ adopting a high mass cut at m$>0.3~M_{\odot}$.
In the scale adopted
here a Salpeter (1955) MF has $\alpha=-2.35$ and a Kroupa (2001) MF would have a
best fitting slope of $\alpha\sim-1.64$.

The estimated slopes cover a wide range from $\alpha=-1.89$ (NGC 6304) to
$\alpha=0.11$ (NGC 2298). Thirteen GCs are in common with the work by Paust et al. (2010),
who estimated MFs using the same photometric dataset and also used multimass dynamical
models. We show the comparison between the two works in Fig. \ref{comp}. The mean 
difference between the two studies is $\Delta \alpha (this~
work-Paust)=0.16\pm0.13$ consistent with only a small (if any) systematic shift. 
However the dispersion about the mean ($\sigma=0.47$) is not 
compatible with the combined errors of the two works. In this context, it should
be noted that the formal error quoted by Paust et al. (2010) as well as those
listed in Table 1 are errors on the MF fit and do not reflect the actual error
budget (due to incomplete radial sampling, errors of the estimated
completeness factor, isochrone/dynamical model inadequacy, etc.). Given the above
considerations, we believe that a more realistic
uncertainty of the MF slopes of both works is of the order of $\sigma_{\alpha}\sim 0.3$.
It is worth noting that for NGC 6093 the difference between the two estimates
exceeds $\Delta \alpha>$1.2. Moreover, for NGC 5466 we find an
unphysical solution with $M_{lum}>M_{dyn}$. Among the GCs in common, these are 
those with the
smallest fraction of stars contained within the ACS field of view. In this
situation, even a small difference in the fitting process can produce large
extrapolation errors. For comparison, a similar analysis of NGC 5466 performed in
Sollima et al. (2017) using MF constraints in the outer portion of
this cluster leads to a significantly flatter MF slope ($\alpha=-0.97$).
The mean difference between the MF slopes derived adopting different low-mass
cuts is $\Delta \alpha_{0.3-0.2}=-0.05\pm0.01$, indicating only a small dependence
of the estimated MF slopes on the adopted lower mass limit.

\begin{figure*}
 \includegraphics[width=14cm]{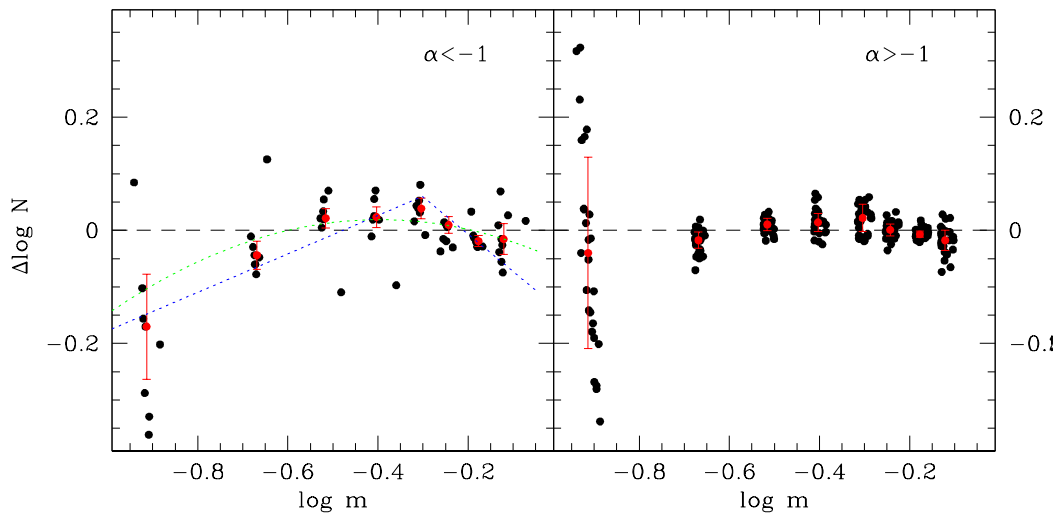}
 \caption{Residuals of the power-law fit for clusters with $\alpha<-1$ (left
 panel) and $\alpha>-1$ (right panel). The average residuals and their standard
 deviations for all mass bin are marked with red dots and errorbars (grey in the
 printed version of the paper). The Kroupa (2001; blue line) and Chabrier (2003;
 green line) IMFs are also marked with dotted lines.}
\label{shape}
\end{figure*}

An inspection of the MFs reveals that while some of them are well fitted by single 
power-laws, others show strong deviations from a power-law MF. To investigate this issue
further, we correlated the $\chi^{2}$ value of the power-law fit with various cluster 
parameters. A single significant correlation has been found with the MF 
slope itself, in the sense that flatter MFs present better power-law fits.
To highlight this result, we plot in Fig. \ref{shape} the residuals of the power-law fit for clusters
with $\alpha\gtrless -1$. It is apparent that while clusters with a relatively
flat MF ($\alpha>-1$) show no significant deviation from the best-fitting
power-law, clusters with steep MFs have a convex shape. In particular, a
point of maximum curvature is apparent at $log (m/M_{\odot})\sim-0.4$
(corresponding to a mass $m\sim0.4~M_{\odot}$).
The same evidence remains apparent even using the $\alpha$ values
calculated adopting a high mass cut at $m>0.3~M_{\odot}$, indicating a 
negligible effect of the uncertainties of the MF estimate at very low masses.

\section{analysis of correlations}
\label{corr_sec}

\begin{figure*}
 \includegraphics[width=14cm]{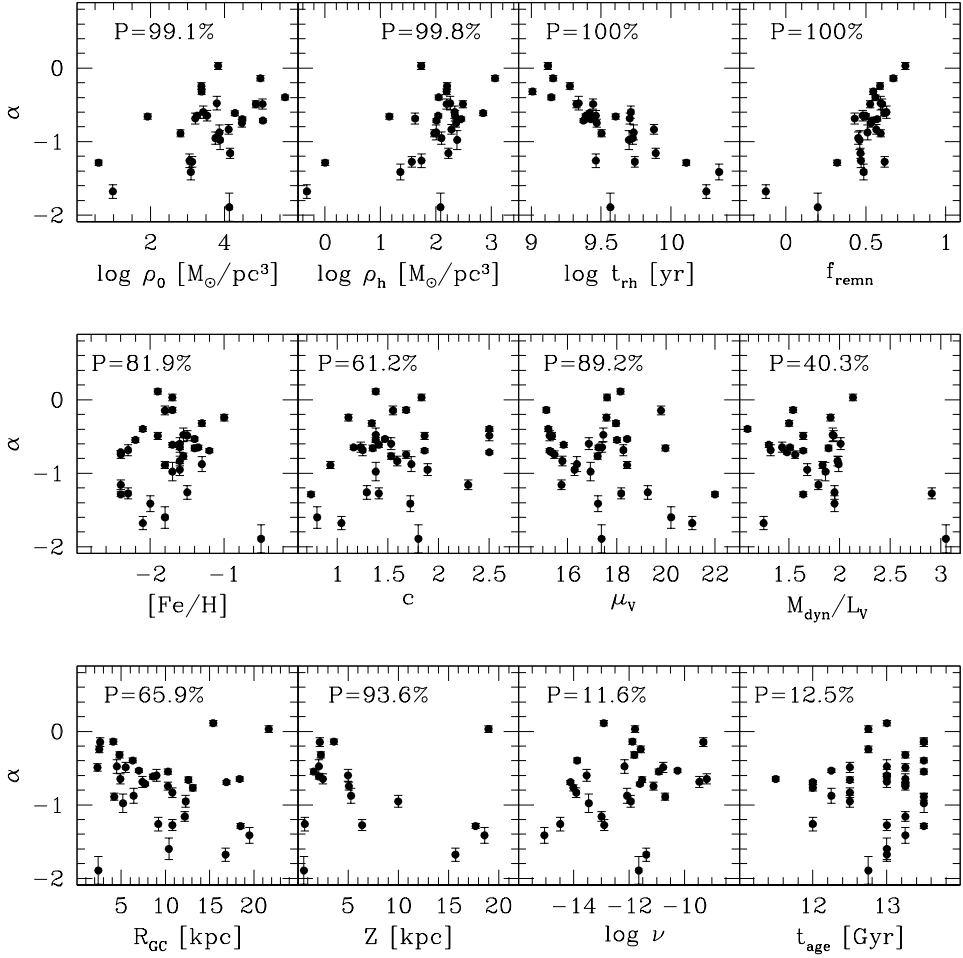}
 \caption{Univariate correlations between the global MF slope $\alpha$ and
 various parameters. The statistical significance (P) of each correlation is
 indicated.}
\label{all}
\end{figure*}

The MF slopes derived here constitute the largest available database and can be
therefore used to search for correlations with other structural and general
parameters.
We considered the following parameters to look for possible correlations with the
MF slope: Position in the Galaxy ($R_{GC},~Z$ from the Harris
catalog; Harris 1996, 2010 edition), destruction rates ($\nu$; from Allen, Moreno \&
Pichardo 2006, 2008), concentration ($c$; from McLaughlin \& van der Marel
2005), age and metallicity ($t_{9},~[Fe/H]$; from Dotter et al. 2010), central
V-band surface brightness, mass-to-light ratio, central and half-mass density, 
half-mass relaxation time and remnant mass fraction
($\mu_{V,0},~M_{dyn}/L_{V},~\rho_{0},~\rho_{h},~t_{rh},~f_{remn}$; from the best fit multimass model
adopted in this analysis).

As a first step, we analysed univariate correlations between $\alpha$ and the
other parameters. For this purpose a Monte Carlo procedure has been applied to
estimate the significance of the obtained correlations. For each of the
considered parameters, we performed an error-weighted least-square fit and 
calculated the $\chi^{2}$ values. Then, the same analysis has been performed on 
one-thousand realizations simulated by randomly swapping
the values of the independent variable among the GCs of the 
sample. The probability that the observed correlation is significant is
therefore given by the fraction of realizations with a $\chi^{2}$ larger than the
observed value. From this approach we found three significant ($P>99.7\%$)
correlations with $log~t_{rh}$, $f_{remn}$ and $log~\rho_{h}$, and a marginally significant
correlation ($95\%<P<99.7\%$) with the central density, while no significant 
correlations have been found with other parameters suggested by previous works
(see Sec.~\ref{intro_sec}). The entire set of correlations and their associated
probabilities are shown in Fig. \ref{all}.

\begin{figure}
 \includegraphics[width=8.6cm]{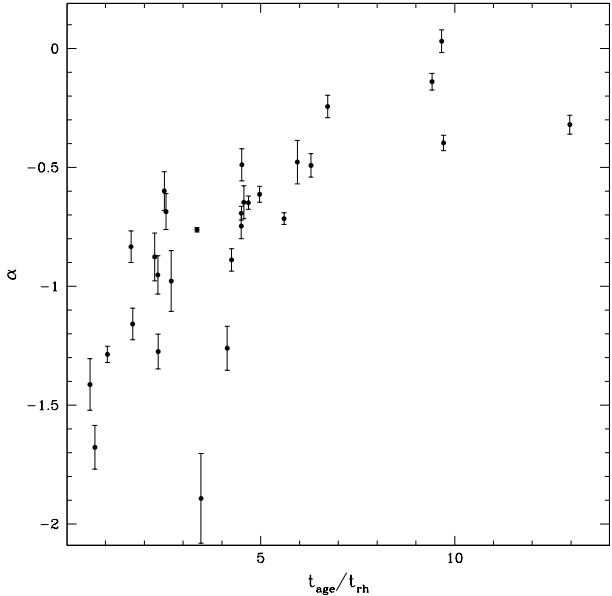}
 \caption{MF slope $\alpha$ as a function of the ratio between cluster age and
 present-day half-mass relaxation time for the 29 GCs of our sample.}
\label{ttrh}
\end{figure}

We note that, the correlation between $\alpha$ and $log~t_{rh}$ has a surprisingly
small dispersion ($\sigma=0.29$ i.e. compatible with the actual $\alpha$
uncertainties; see Sec.~\ref{res_sec}).
The only cluster straying from the observed trend is NGC 6304 with a 
MF slope $\alpha=-1.89\pm0.19$ steeper than those of the other GCs
of our sample. It should be noted however that this cluster is close to the
Galactic bulge whose MF is known to be bottom-heavy (Zoccali et al. 2000; 
Calamida et al. 2015). We therefore cannot exclude the possibility that the peculiar 
MF measured in this GC is due to a significant contamination from the bulge.
Another way to visualize the above correlation is shown in Fig. \ref{ttrh} where
the MF slope $\alpha$ is plotted against the ratio of age to 
present-day half-mass relaxation time $t_{age}/t_{rh}$. Again, GCs define a very tight
correlation in this plot, indicating an evolutionary sequence. In other words, after the
same number of present-day half-mass relaxation times clusters have similar MF
slopes regardless of their orbits and chemical compositions.
Another investigation can be made based on the location of clusters in the $t_{age}/t_{rh}-\alpha$ 
plane as shown in Fig. \ref{ttrh}: while GCs with $t_{age}/t_{rh}<1$ have MF slopes $\alpha\sim-1.5$
similar to that of a Kroupa (2001) MF, the mean MF slope increases by $\sim$0.5
dex at $t_{age}/t_{rh}\sim3$.

We also extended our analysis to bivariate correlations.
The $\chi^{2}$ of a bi-linear fit of all the possible pairs of parameters has
been calculated and compared. The smallest $\chi^{2}$ are all those found by 
assuming $log~t_{rh}$ as the independent variable. To estimate how significant
the improvement with respect to an univariate fit is, we applied the same Monte
Carlo approach described above: we compared the $\chi^{2}$ of the bivariate fit
(assuming $log~t_{rh}$ as the first independent variable)
with those obtained by randomly swapping the values of the adopted 
second independent variable.
We found a marginally significant (P=97.7\%) improvement with respect to the
univariate fit only by assuming as second independent variable $f_{remn}$.

\begin{figure*}
 \includegraphics[width=14cm]{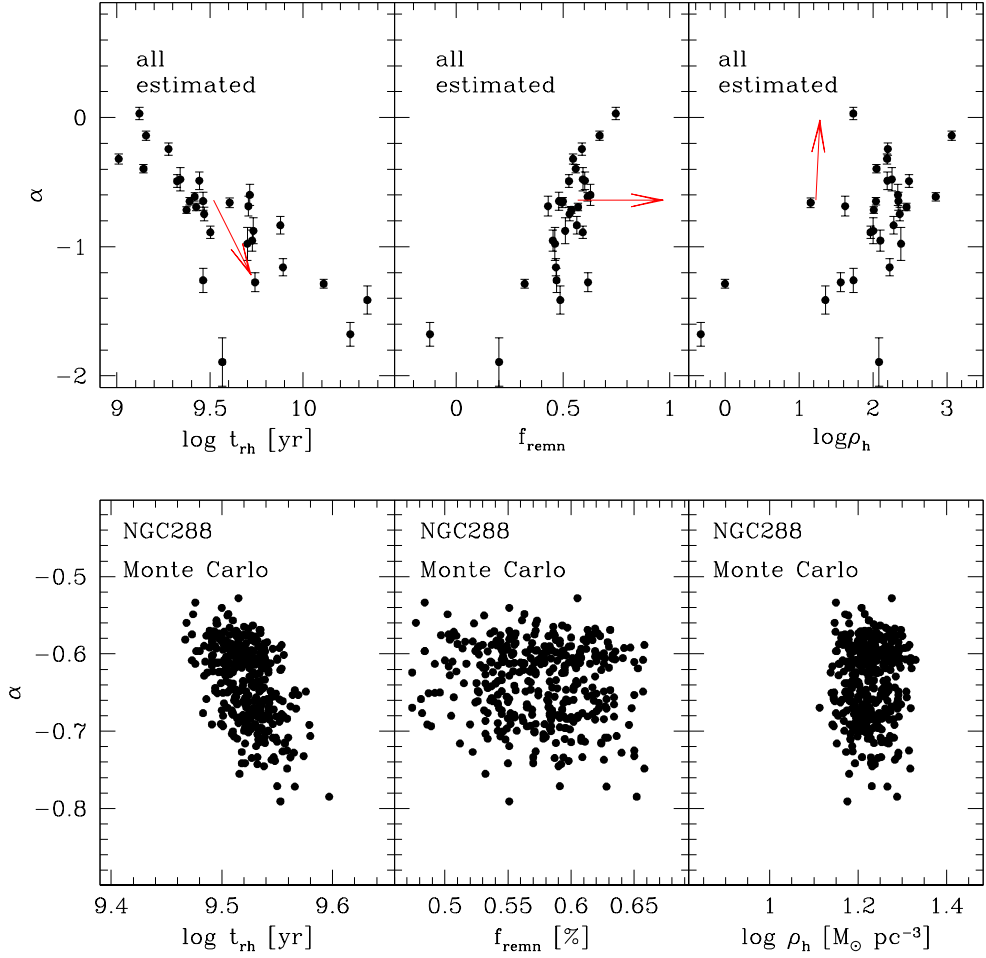}
 \caption{Bottom panels: distribution of the accepted trials of the 
 Markov-Chain Monte Carlo in the $log~t_{rh}-\alpha$ (left), $f_{remn}-\alpha$
 (middle) and $log~\rho_{h}-\alpha$ (right) planes. Top panels: distribution 
 of the 29 analysed clusters in the $log~t_{rh}-\alpha$ (left), $f_{remn}-\alpha$
 (middle) and $log~\rho_{h}-\alpha$
 (right) planes. The orientation of the biases are indicated by arrows.}
\label{monte}
\end{figure*}

Unfortunately, $t_{rh}$, $f_{remn}$, $\rho_{h}$ and $\alpha$ are all output parameters of
the best fit multimass model adopted in this work. It is therefore
possible that covariances between the uncertainties in these parameters
 conspire to spuriously create the quoted correlations. To test this
hypothesis, we run a Markov-Chain Monte Carlo analysis on our target GCs. This
algorithm samples the parameter space in the neighbourhood of the best fit
parameters, providing a distribution of accepted trials which reflect the actual
probability distribution. The distribution of accepted trial values in the
$log~t_{rh}-\alpha$, $f_{remn}-\alpha$ and $log~\rho_{h}-\alpha$ planes for the 
cluster NGC 288 are shown in 
Fig. \ref{monte}, as an example. The covariance between $\alpha$ and $log~t_{rh}$ 
is apparent with a tendency of solutions with longer $t_{rh}$ to have 
shallower MFs, while no significant slope with $log~\rho_{h}$ or $f_{remn}$ is noticeable. 
A similar behaviour has been noticed in the other clusters with no significant
dependence of the bias orientation and strength on the position in these planes.
Note that the direction of such a bias in the $log~t_{rh}-\alpha$ plane is similar 
to that of the observed correlation.
However, the shift in $\alpha$ along the bias direction needed to erase any
significant correlation with $log~t_{rh}$ 
would be as large as $\Delta \alpha\sim0.8$ at the extreme of this plot i.e.
$\sim 3$ times larger than typical uncertainties.
To further check that the observed correlation is not driven by the
covariance spuriously introduced by the adopted fitting procedure, we correlated
the derived MF slopes with the $log~t_{rh}$ values derived independently by
McLaughlin \& van der Marel (2005) fitting single mass King (1966) models. 
Also in this case, we found a confidence level $>$99.9\% that the two variables are correlated.
So, while it is conceivable that the
observed $log~t_{rh}-\alpha$ correlation is sharpened by the covariance between 
errors, it cannot be completely produced by this bias.  

Another source of bias could be linked to an overestimate of the
level of completeness. This could indeed spuriously deplete the MF at its
low-mass end, in particular for dense GCs characterized by short relaxation
times. While we cannot completely exclude this occurrence, it is unlikely that a
significant bias in the estimated completeness is present above the
magnitude limit corresponding to stellar masses $m>0.2~M_{\odot}$, in the 
portion of the CMD used to estimate the MF slope. 
To check the possible effect of uncertainties in the completeness correction we
repeated the analysis by using the MF slope $\alpha$ calculated assuming a high
$m>0.3~M_{\odot}$ cut and excluding all those clusters with completeness levels
$<$50\% within their core radii at masses $<0.25~M_{\odot}$ (see Leigh et al.
2012). Although only 15 GCs survive to this severe criterion, the correlation
between $\alpha$ and $log~t_{rh}$ remains significant at $99.9\%$, while the
significance levels of the other correlations drop below 75\%. On the basis of
the above test, we conclude that the $log~t_{rh}-\alpha$ correlation we observe 
among the GCs of our sample is robust and real.

\section{Discussion}

Through a comparison of the deepest available HST photometric data with
multimass King-Michie models we derived the MFs of 35 Galactic GCs
just above the
hydrogen burning limit as well as structural parameters, masses, M/L ratios and fraction of
remnants for a subset of 29 GCs with available radial velocity information.

The MFs of GCs are generally well described by power-laws, in particular when
clusters with relatively shallow MF slopes ($\alpha>-1$) are analysed. 
Noticeable deviations from
power-laws are evident in clusters whose MFs are steeper than $\alpha=-1$.
In particular, in these cases a bend in the MF is appreciable at masses
$m\sim0.4~M_{\odot}$ with a significant depletion of low-mass stars.
This evidence has been previously reported by De Marchi \& Paresce (1997) and
De Marchi, Paresce \& Portegies-Zwart (2010) who also defined a relation linking
the position of the bend with the dissolution timescale.
These GCs have half-mass relaxation times of $t_{rh}>6$ Gyr, and it is
possible that their MFs have not been significantly altered by dynamical 
evolution effects. We will discuss these deviations in greater detail
in a second paper (Baumgardt \& Sollima 2017) where we compare the MF slopes derived in this work
with the results of $N$-body simulations. 
In this interpretation, the convex shape of their MFs might
resemble the original shape of the IMF, in agreement with the prediction of
Kroupa (2001) and Chabrier (2003).
 
By correlating the derived MF slopes with different structural and orbital
cluster parameters we found significant and tight correlations with the
half-mass relaxation time.
Although covariance between uncertainties can spuriously enhance the strength of
the $log~t_{rh}-\alpha$ correlation, we believe that it is real since
{\it i)} its extent exceeds the range over which the above mentioned bias would
produce a sizeable effect, and {\it
ii)} it is significant using independent estimates of $t_{rh}$.
This correlation is expected as a result of the natural evolution of
collisional stellar systems. Indeed, two-body relaxation is the main mechanism
leading to the segregation of low-mass stars to the outer cluster parts, where they
can be easily lost by tidal stripping (Vesperini \&
Heggie 1997; Baumgardt \& Makino 2003; Leigh et al. 2012). So, the shorter
the timescale of internal dynamical evolution the more efficient is the
depletion of the MF. 
However, the location of N-body simulations in the $log~t_{rh}-\alpha$ plane is
highly sensitive to the original slope of the IMF, with clusters starting 
with a steeper IMF reaching also steeper present-day MFs
 after a given number of elapsed relaxation times 
than clusters starting with flatter IMFs (Webb \& Vesperini 2016). Thus, a significant
spread in this relation would be apparent if cluster-to-cluster
variations of the IMF were present at the epoch of their formation.
On the other hand, all surveys of N-body simulations performed so far,
showed that two-body relaxation is expected to produce a slow variation of the 
MF slope. In particular, simulations starting with a Kroupa (2001) IMF take 
$\sim13$ half-mass relaxation times to flatten their MF up to a slope of
$\alpha=-1$ and reach a flat $\alpha=0$ slope only close to dissolution
(Baumgardt \& Makino 2003; Webb \& Vesperini 2014; Lamers et al. 2013). In such a picture, it is hard
to explain the large range in $\alpha$ covered by the GCs of our sample, in
particular in the less evolved $t_{age}/t_{rh}<5$ regime, without 
any primordial spread in their IMFs. In this last case, however, a correlation
between the IMF slope and the present-day half-mass relaxation time would be 
necessary to reproduce the observed $log~t_{rh}-\alpha$ correlation.

The universality of the IMF of Milky Way GCs has
important implications for the thermodynamics of the gas clouds from which
GCs formed at high-redshift. 
Theoretical arguments indeed suggest a dependence of the IMF on the metal content and
the initial density of the cluster because of their effect on the Jeans mass and
on the efficiency of radiative feedback (Silk 1977; Adams \& Fatuzzo 1996; 
Larson 1998; Klessen, Spaans \& Jappsen 2007) although the actual impact of
these processes is uncertain. Whether Milky Way GCs were born with a Universal
IMF or not, provides insight on the efficiency of these mechanisms in the 
environmental conditions of GCs at their birth.
It is interesting to consider the evidence found in stellar systems
populating regions of the $M_{V}-r_{eff}$ plane contiguous to GCs. 
In this regard, while Grillmair et al. (1998) and Wyse et al. (2002) derived MFs 
for Draco and Ursa Minor dwarf spheroidals which are consistent with a Salpeter (1955)
IMF, Geha et al. (2013) found evidence of MF
variations correlated with the mean metallicity in a sample of ultra-faint dwarf
galaxies. Since these systems are dynamically unevolved, these variations
can be only interpreted as primordial. This study has been however questioned by
El-Badry, Weisz \& Quataert (2017) who found that significant MF differences
cannot be detected unless the photometric data used is significantly deeper than 
that currently available.
On the other hand, Weisz et al. (2013) 
analysed a large sample of young clusters and
associations whose MFs are available in the literature. In spite of the large cluster-to-cluster differences, a careful
revision of the associated errors indicates that the hypothesis that they are
consistent with a single IMF slope cannot be ruled out. Hence, due to the above
conflicting results, it is not clear if a common mechanism of star formation was
at work for GCs and less massive and dense stellar systems.

Interestingly, Baumgardt \& Makino (2003), Lamers et al (2013) and Webb \& Leigh (2015) found a unique relation linking the
present-day MF slope and the fraction of mass lost by their simulated GCs. Such
a relation, which is valid only if a universal IMF is assumed, appears to be
almost insensitive to the strength of the tidal field, the type of cluster
orbit, and to the initial mass and size of the cluster.
The MFs derived here have slopes which imply a huge amount of mass lost 
($>70\%$) by the majority of GCs in our sample.
By inverting eq. 14 of Baumgardt \& Makino (2003) and adopting the present-day
masses listed in Table 1, we estimated the
amount of mass lost by each cluster during its evolution. Assuming 
that our sample covers $\sim$20\% of the GC system of the Milky Way, a global
mass of $\sim2\times10^{8}~M_{\odot}$, mainly in low-mass stars, could have been
released in the Galactic halo by GCs. In spite of the large uncertainties in the
Galactic halo mass (Morrison 1993; Bell et al. 2008; Deason, Belokurov \& Evans 2011), this 
could constitute a significant contribution to the
total mass budget of the halo. This is in agreement with the prediction by
Martell \& Grebel (2010) based on the fraction of halo stars showing the 
chemical signature of GCs stars. In this picture, one would expect a significant
excess of low-mass stars in the MF of halo stars. Such a prediction could be
probably verified by the incoming data provided by the GAIA survey.

Another significant correlation has been found between the present-day MF slopes
and the fraction of dark mass. The natural interpretation of this correlation is
that dark remnants (mainly constituted by white dwarfs) have
masses larger than the average cluster stars and are being more efficiently retained.
Moreover, the fraction of white dwarfs steadily increase with time as less massive (and
more abundant) stars approach this late stage of their evolution. Studies based
on N-body simulations have shown that, because of the two above mentioned
processes, the fraction of mass contained in remnants increases as two-body
relaxation proceeds (Baumgardt \& Makino 2003; Contenta, Varri \& Heggie 2015). 
Such a correlation becomes less significant
when dense GCs, subject to low completeness at low masses are excluded.

Significant correlations with the half-mass and (marginally) with the central density (obviously related
to $t_{rh}$) have also been found, in agreement with previous finding by Paust et
al. (2010). All the correlations with orbital parameters and position in the
Galaxy suggested by previous studies based on the analysis of small samples of 
GCs (Capaccioli et al. 1993; Djorgovski et al. 1993; Piotto \& Zoccali 1999), 
have been found to be less significant although they cannot be completely 
ruled out. No significant correlation has been found with the cluster
concentration, as previously suggested by De Marchi et al. (2007). Note that,
while the uncertainties on the individual MF slopes do not allow to exclude the 
presence of such a correlation, for most GCs studied by De Marchi et al. (2007) 
the slope of the global MF has been assumed to be that measured in an external
region close to the half-mass radius. However, in high-concentration clusters
the analysed fields are often located well beyond the half-mass radii estimated here,
a region where the MF is expected to be steeper because of mass segregation 
effects. In particular, there are three high concentration GCs out of 6 in 
their sample where the MF is calculated between 3 and 7 half-mass radii. 
This could create a spurious correlation between MF slope and concentration. 

It is worth stressing that our results are based on an analysis conducted in the 
central region of GCs where
mass segregation effects are particularly strong. As a consequence, the derived
global MFs are sensitive to the recipe of mass segregation of the adopted
multimass models. In Sollima et al. (2015) we showed that such an assumption
can potentially lead to biases in the estimated MF slopes as large as $\Delta
\alpha\sim0.2$, i.e. comparable with the estimated random uncertainties
(see Sect. \ref{res_sec}). Although the magnitude of such a bias cannot alter the conclusions 
of this paper,
the present analysis would greatly benefit from constraints on the MF measured in
the outer regions of these clusters (see e.g. Sollima et al. 2017). 

\section*{Acknowledgments}

We warmly thank Michele Bellazzini, Enrico Vesperini and Luca Ciotti for useful discussions.
We also thank the anonymous referee for his/her helpful comments and
suggestions.

\label{lastpage}


\begin{thebibliography}{99}

\bibitem[\protect\citeauthoryear{Adams \& Fatuzzo}{1996}]{1996ApJ...464..256A} Adams F.~C., Fatuzzo M., 1996, ApJ, 464, 256 
\bibitem[\protect\citeauthoryear{Allen, Moreno, \& Pichardo}{2006}]{2006ApJ...652.1150A} Allen C., Moreno E., Pichardo B., 2006, ApJ, 652, 1150 
\bibitem[\protect\citeauthoryear{Allen, Moreno, \& Pichardo}{2008}]{2008ApJ...674..237A} Allen C., Moreno E., Pichardo B., 2008, ApJ, 674, 237-246 
\bibitem[\protect\citeauthoryear{Alonso-Garc{\'{\i}}a et al.}{2012}]{2012AJ....143...70A} Alonso-Garc{\'{\i}}a J., Mateo M., Sen B., Banerjee M., Catelan M., Minniti D., von Braun K., 2012, AJ, 143, 70 
\bibitem[\protect\citeauthoryear{Anderson et al.}{2008}]{2008AJ....135.2055A} Anderson J., et al., 2008, AJ, 135, 2055 
\bibitem[\protect\citeauthoryear{Bastian, Covey, \& Meyer}{2010}]{2010ARA&A..48..339B} Bastian N., Covey K.~R., Meyer M.~R., 2010, ARA\&A, 48, 339 
\bibitem[\protect\citeauthoryear{Baumgardt}{2017}]{2017MNRAS.464.2174B} Baumgardt H., 2017, MNRAS, 464, 2174 
\bibitem[\protect\citeauthoryear{Baumgardt \& Makino}{2003}]{2003MNRAS.340..227B} Baumgardt H., Makino J., 2003, MNRAS, 340, 227 
\bibitem[\protect\citeauthoryear{Bell et al.}{2008}]{2008ApJ...680..295B} Bell E.~F., et al., 2008, ApJ, 680, 295-311 
\bibitem[\protect\citeauthoryear{Bonnell et al.}{1997}]{1997MNRAS.285..201B} Bonnell I.~A., Bate M.~R., Clarke C.~J., Pringle J.~E., 1997, MNRAS, 285, 201 
\bibitem[\protect\citeauthoryear{Calamida et al.}{2015}]{2015ApJ...810....8C} Calamida A., et al., 2015, ApJ, 810, 8 
\bibitem[\protect\citeauthoryear{Capaccioli, Piotto, \& Stiavelli}{1993}]{1993MNRAS.261..819C} Capaccioli M., Piotto G., Stiavelli M., 1993, MNRAS, 261, 819 
\bibitem[\protect\citeauthoryear{Chabrier}{2003}]{2003PASP..115..763C} Chabrier G., 2003, PASP, 115, 763 
\bibitem[\protect\citeauthoryear{Chabrier \& Mera}{1997}]{1997A&A...328...83C} Chabrier G., Mera D., 1997, A\&A, 328, 83 
\bibitem[\protect\citeauthoryear{Contenta, Varri, \& Heggie}{2015}]{2015MNRAS.449L.100C} Contenta F., Varri A.~L., Heggie D.~C., 2015, MNRAS, 449, L100 
\bibitem[\protect\citeauthoryear{Da Costa}{1982}]{1982AJ.....87..990D} Da Costa G.~S., 1982, AJ, 87, 990 
\bibitem[\protect\citeauthoryear{Deason, Belokurov, \& Evans}{2011}]{2011MNRAS.416.2903D} Deason A.~J., Belokurov V., Evans N.~W., 2011, MNRAS, 416, 2903 
\bibitem[\protect\citeauthoryear{De Marchi \& Paresce}{1997}]{1997ApJ...476L..19D} De Marchi G., Paresce F., 1997, ApJ, 476, L19 
\bibitem[\protect\citeauthoryear{De Marchi, Paresce, \& Portegies Zwart}{2010}]{2010ApJ...718..105D} De Marchi G., Paresce F., Portegies Zwart S., 2010, ApJ, 718, 105 
\bibitem[\protect\citeauthoryear{De Marchi, Paresce, \& Pulone}{2007}]{2007ApJ...656L..65D} De Marchi G., Paresce F., Pulone L., 2007, ApJ, 656, L65 
\bibitem[\protect\citeauthoryear{Djorgovski \& King}{1984}]{1984ApJ...277L..49D} Djorgovski S., King I.~R., 1984, ApJ, 277, L49 
\bibitem[\protect\citeauthoryear{Djorgovski, Piotto, \& Capaccioli}{1993}]{1993AJ....105.2148D} Djorgovski S., Piotto G., Capaccioli M., 1993, AJ, 105, 2148 
\bibitem[\protect\citeauthoryear{Dotter et al.}{2010}]{2010ApJ...708..698D} Dotter A., et al., 2010, ApJ, 708, 698 
\bibitem[\protect\citeauthoryear{Dotter et al.}{2007}]{2007AJ....134..376D} Dotter A., Chaboyer B., Jevremovi{\'c} D., Baron E., Ferguson J.~W., Sarajedini A., Anderson J., 2007, AJ, 134, 376 
\bibitem[\protect\citeauthoryear{El-Badry, Weisz, \& Quataert}{2017}]{2017MNRAS.468..319E} El-Badry K., Weisz D.~R., Quataert E., 2017, MNRAS, 468, 319 
\bibitem[\protect\citeauthoryear{Fleck}{1982}]{1982MNRAS...201...551F} Fleck
R.~C,, Jr., 1982, MNRAS, 201, 551 
\bibitem[\protect\citeauthoryear{Geha et al.}{2013}]{2013ApJ...771...29G} Geha M., et al., 2013, ApJ, 771, 29 
\bibitem[\protect\citeauthoryear{Giersz \&
Heggie}{1996}]{1996MNRAS...279...1037G} Giersz M., Heggie D.~C., 1996, MNRAS,
279, 1037 
\bibitem[\protect\citeauthoryear{Grillmair et al.}{1998}]{1998AJ....115..144G} Grillmair C.~J., et al., 1998, AJ, 115, 144 
\bibitem[\protect\citeauthoryear{Gunn 
\& Griffin}{1979}]{1979AJ.....84..752G} Gunn J.~E., Griffin R.~F., 1979, AJ, 84, 752 
\bibitem[\protect\citeauthoryear{Harris}{1986}]{1996AJ....112..1487H} Harris W.~E., 1996, AJ, 112, 1487
\bibitem[\protect\citeauthoryear{Kalirai et al.}{2009}]{2009ApJ...705..408K} Kalirai J.~S., Saul Davis D., Richer H.~B., Bergeron P., Catelan M., Hansen B.~M.~S., Rich R.~M., 2009, ApJ, 705, 408 
\bibitem[\protect\citeauthoryear{King}{1966}]{1966AJ.....71...64K} King 
I.~R., 1966, AJ, 71, 64 
\bibitem[\protect\citeauthoryear{Klessen, Spaans, \& Jappsen}{2007}]{2007MNRAS.374L..29K} Klessen R.~S., Spaans M., Jappsen A.-K., 2007, MNRAS, 374, L29 
\bibitem[\protect\citeauthoryear{Kroupa}{2001}]{2001MNRAS.322..231K} Kroupa 
P., 2001, MNRAS, 322, 231 
\bibitem[\protect\citeauthoryear{Kruijssen 
\& Mieske}{2009}]{2009A&A...500..785K} Kruijssen J.~M.~D., Mieske S., 2009, A\&A, 500, 785 
\bibitem[\protect\citeauthoryear{Lamers, Baumgardt, \& Gieles}{2013}]{2013MNRAS.433.1378L} Lamers H.~J.~G.~L.~M., Baumgardt H., Gieles M., 2013, MNRAS, 433, 1378 
\bibitem[\protect\citeauthoryear{Larson}{1998}]{1998MNRAS.301..569L} Larson R.~B., 1998, MNRAS, 301, 569 
\bibitem[\protect\citeauthoryear{Leigh et al.}{2012}]{2012MNRAS.422.1592L} Leigh N., Umbreit S., Sills A., Knigge C., de Marchi G., Glebbeek E., Sarajedini A., 2012, MNRAS, 422, 1592 
\bibitem[\protect\citeauthoryear{Martell \& Grebel}{2010}]{2010A&A...519A..14M} Martell S.~L., Grebel E.~K., 2010, A\&A, 519, A14 
\bibitem[\protect\citeauthoryear{Melbourne et al.}{2000}]{2000AJ....120.3127M} Melbourne J., Sarajedini A., Layden A., Martins D.~H., 2000, AJ, 120, 3127 
\bibitem[\protect\citeauthoryear{Miller \& Scalo}{1979}]{1979ApJS...41..513M} Miller G.~E., Scalo J.~M., 1979, ApJS, 41, 513 
\bibitem[\protect\citeauthoryear{Miocchi et al.}{2013}]{2013ApJ...774..151M} Miocchi P., et al., 2013, ApJ, 774, 151 
\bibitem[\protect\citeauthoryear{McClure et al.}{1986}]{1986ApJ...307L..49M} McClure R.~D., et al., 1986, ApJ, 307, L49 
\bibitem[\protect\citeauthoryear{McLaughlin 
\& van der Marel}{2005}]{2005ApJS..161..304M} McLaughlin D.~E., van der Marel R.~P., 2005, ApJS, 161, 304 
\bibitem[\protect\citeauthoryear{Morrison}{1993}]{1993AJ....106..578M} Morrison H.~L., 1993, AJ, 106, 578 
\bibitem[\protect\citeauthoryear{Nakamura \& Umemura}{2001}]{2001ApJ...548...19N} Nakamura F., Umemura M., 2001, ApJ, 548, 19 
\bibitem[\protect\citeauthoryear{Paresce \& De Marchi}{2000}]{2000ApJ...534..870P} Paresce F., De Marchi G., 2000, ApJ, 534, 870 
\bibitem[\protect\citeauthoryear{Paust, Wilson, \& van Belle}{2014}]{2014AJ....148...19P} Paust N., Wilson D., van Belle G., 2014, AJ, 148, 19 
\bibitem[\protect\citeauthoryear{Paust et al.}{2010}]{2010AJ....139..476P} Paust N.~E.~Q., et al., 2010, AJ, 139, 476 
\bibitem[\protect\citeauthoryear{Piotto \& Zoccali}{1999}]{1999A&A...345..485P} Piotto G., Zoccali M., 1999, A\&A, 345, 485 
\bibitem[\protect\citeauthoryear{Pulone et al.}{2003}]{2003A&A...399..121P} Pulone L., De Marchi G., Covino S., Paresce F., 2003, A\&A, 399, 121 
\bibitem[\protect\citeauthoryear{Richer et al.}{1990}]{1990ApJ...359L..11R} Richer H.~B., Fahlman G.~G., Buonanno R., Fusi Pecci F., 1990, ApJ, 359, L11 
\bibitem[\protect\citeauthoryear{Robin et al.}{2003}]{2003A&A...409..523R} Robin A.~C., Reyl{\'e} C., Derri{\`e}re S., Picaud S., 2003, A\&A, 409, 523 
\bibitem[\protect\citeauthoryear{Salpeter}{1955}]{1955ApJ...121..161S} Salpeter E.~E., 1955, ApJ, 121, 161 
\bibitem[\protect\citeauthoryear{Santiago, Elson, \& Gilmore}{1996}]{1996MNRAS.281.1363S} Santiago B.~X., Elson R.~A.~W., Gilmore G.~F., 1996, MNRAS, 281, 1363 
\bibitem[\protect\citeauthoryear{Sarajedini et 
al.}{2007}]{2007AJ....133.1658S} Sarajedini A., et al., 2007, AJ, 133, 1658 
\bibitem[\protect\citeauthoryear{Silk}{1977}]{1977ApJ...214..718S} Silk J., 1977, ApJ, 214, 718 
\bibitem[\protect\citeauthoryear{Sollima, Bellazzini, 
\& Lee}{2012}]{2012ApJ...755..156S} Sollima A., Bellazzini M., Lee J.-W., 2012, ApJ, 755, 156 
\bibitem[\protect\citeauthoryear{Sollima et al.}{2015}]{2015MNRAS.451.2185S} Sollima A., Baumgardt H., Zocchi A., Balbinot E., Gieles M., H{\'e}nault-Brunet V., Varri A.~L., 2015, MNRAS, 451, 2185 
\bibitem[\protect\citeauthoryear{Sollima et al.}{2017}]{2017MNRAS.464.3871S} Sollima A., Dalessandro E., Beccari G., Pallanca C., 2017, MNRAS, 464, 3871 
\bibitem[\protect\citeauthoryear{Spitzer}{1987}]{1987degc.book.....S} Spitzer
L., 1987, "Dynamical evolution of globular clusters", Princeton Univ. Press,
Princeton
\bibitem[\protect\citeauthoryear{Trager, King, \& Djorgovski}{1995}]{1995AJ....109..218T} Trager S.~C., King I.~R., Djorgovski S., 1995, AJ, 109, 218 
\bibitem[\protect\citeauthoryear{Vesperini \& Heggie}{1997}]{1997MNRAS.289..898V} Vesperini E., Heggie D.~C., 1997, MNRAS, 289, 898 
\bibitem[\protect\citeauthoryear{Watkins et al.}{2015}]{2015ApJ...803...29W} Watkins L.~L., van der Marel R.~P., Bellini A., Anderson J., 2015, ApJ, 803, 29 
\bibitem[\protect\citeauthoryear{Webb \& Leigh}{2015}]{2015MNRAS.453.3278W} Webb
J.~J., Leigh N. W. C., 2015, MNRAS, 453, 3278 
\bibitem[\protect\citeauthoryear{Webb \& Vesperini}{2016}]{2016MNRAS.463.2383W} Webb J.~J., Vesperini E., 2016, MNRAS, 463, 2383 
\bibitem[\protect\citeauthoryear{Weisz et al.}{2013}]{2013ApJ...762..123W} Weisz D.~R., et al., 2013, ApJ, 762, 123 
\bibitem[\protect\citeauthoryear{Wyse et al.}{2002}]{2002NewA....7..395W} Wyse R.~F.~G., Gilmore G., Houdashelt M.~L., Feltzing S., Hebb L., Gallagher J.~S., III, Smecker-Hane T.~A., 2002, NewA, 7, 395 
\bibitem[\protect\citeauthoryear{Zoccali et al.}{2000}]{2000ApJ...530..418Z} Zoccali M., Cassisi S., Frogel J.~A., Gould A., Ortolani S., Renzini A., Rich R.~M., Stephens A.~W., 2000, ApJ, 530, 418 

\end{thebibliography}
\end{document}